\documentclass[showpacs,showkeys]{revtex4}

\usepackage{amssymb}
\usepackage{amsmath}
\usepackage{amsfonts}
\usepackage{graphicx}
\providecommand{\rbr}[1]{\left( #1 \right)} %
\providecommand{\sqbr}[1]{\left[ #1 \right]} %
\providecommand{\mt}[1]{\mathrm{#1}} %
\providecommand{\mc}[1]{\mathcal{#1}}%
\providecommand{\mb}[1]{\mathbb{#1}}%
\providecommand{\tprod}[1]{\sideset{}{_{\otimes_{#1}}}\prod}%
\def\ra{\rightarrow}
\def\lra{\longrightarrow}
\def\BG{S_{\mathrm{BG}}}
\def\bxi{\boldsymbol\xi}
\def\bxi{\boldsymbol\xi}

\begin{document}
\title[ ]{Complete versus incomplete definitions of the deformed logarithmic and exponential functions}
\author{Thomas Oikonomou}

\email{thoikonomou@chem.demokritos.gr}

\affiliation{Centro Brasileiro de Pesquisas Fisicas, Xavier Sigaud
150, 22290-180 Rio de Janeiro-RJ, Brazil\\ and \\ Institute of
Physical Chemistry, National Center for Scientific Research
``Demokritos", 15310 Athens, Greece}

\author{G. Baris Bagci}

\affiliation{Department of Physics, Faculty of Science, Ege
University, 35100 Izmir, Turkey}

\keywords{deformed logarithm/exponential; deformation parameter
range; Tsallis/Kaniadakis statistics; Abe/Borges-Roditi deformed
functions}
\pacs{02.50.-r; 05.20.-y; 05.20.-Gg; 05.90.+m}

\begin{abstract}
The recent generalizations of Boltzmann-Gibbs statistics
mathematically relies on the deformed logarithmic and exponential
functions defined through some deformation parameters. In the
present work, we investigate whether a deformed
logarithmic/exponential map is a bijection from
$\mathbb{R}^+/\mathbb{R}$ (set of positive real numbers/all real
numbers) to $\mathbb{R}/\mathbb{R}^+$, as their undeformed
counterparts. We show that their inverse map exists only in some
subsets of the aforementioned (co)domains. Furthermore, we present
conditions which a generalized deformed function has to satisfy, so
that the most important properties of the ordinary functions are
preserved. The fulfillment of these conditions permits us to
determine the validity interval of the deformation parameters. We
finally apply our analysis to Tsallis, Kaniadakis, Abe and
Borges-Roditi deformed functions.
\end{abstract}
\eid{ }
\date{\today }
\startpage{1}
\endpage{1}
\maketitle

\section{Introduction}\label{intro}
%
In order to generalize the Boltzmann--Gibbs (BG) entropy and associated statistics, there have been a multitude of entropy definitions such as Tsallis \cite{Tsallis1988}, Kaniadakis \cite{Kaniadakis02}, Borges-Roditi (BR) \cite{BorgesRoditi}, Abe \cite{Abe97}  and R\'enyi \cite{Renyi1970} generalized entropies. Central goal of these generalizations is the derivation of
non-exponential distributions, observed in a large number of
phenomena in various fields of physical sciences involving, among
others, long-range correlations, non-ergodicity, long-time memories,
(multi)fractal structures and self-similarities. An important
thermostatistics in this regard is the one proposed originally by
Tsallis ($q$-statistics) \cite{Tsallis1988}, yielding asymptotically
inverse power law distributions. The $q$-statistics has been applied
to a vast number of physical situations in various scientific areas,
of which a comprehensive review can be found in Ref. \cite{Tsnewbook}. Similarly, one can obtain power law distributions
in the thermodynamic limit through a different generalization
introduced by Kaniadakis ($\kappa$-statistics)
\cite{Kaniadakis02,KaniadQuaratiScarf,Kaniadakis05}, developed in
the context of special relativity.

Considering trace-form entropies, $\sum_i\Lambda_{\bxi}(p_i)=$
$\sum_ip_i\frac{\Lambda_{\bxi}(p_i)}{p_i}$, where
$\bxi=\{\xi_j\}_{j=1,\ldots,u}$ is a deformation parameter set,
$p_i$ is the probability of the $i$th-microstate and
$\Lambda_{\bxi}(p_i)$ is a probability functional, the
generalization procedure consists of introducing deformed
logarithmic and exponential functions \cite{Oik-Tirn2008},
$f_{\bxi}:\mb{F}^+\subseteq \mb{R}^+\ra\mb{F}\subseteq \mb{R}$ with
$f_{\bxi}(x):=x\,\Lambda_{\bxi}(1/x)$ where
$\mb{F}\,(\mb{F}^{+})$ is a subset
of all (positive) real numbers. Similarly, the deformed inverse function
$f^{-1}_{\bxi}$ is a mapping, namely,
$f^{-1}_{\bxi}:\mb{F}\subseteq\mb{R}\ra\mb{F}^+\subseteq\mb{R}^+$
with $f^{-1}_{\bxi}(x):=\mt{Inverse}\{x\,\Lambda_{\bxi}(1/x)\}$.
These deformed functions are required to recover the ordinary
definitions for specific (and unique) values $\bxi_0$ of the
deformation parameters i.e., $f_{\bxi_0}(x)=\ln(x)$ and
$f^{-1}_{\bxi_0}(x)=\exp(x)$. The aforementioned $\bxi$--generalized
functions usually present mathematical properties which differ
from the ordinary, undeformed  ones. For example, we generally
observe for the deformed logarithms
%
\begin{align}\label{eq:1-logq}
f_{\bxi}(x\, y)&\ne f_{\bxi}(x)+f_{\bxi}(y)
\end{align}
%
when $\bxi\ne\bxi_0$. Equivalently, assuming the existence of the
inverse function $f^{-1}_{\bxi}(x)$, the equation above can be
written in terms of the deformed exponentials as
\begin{align}\label{eq:2-expq}
f^{-1}_{\bxi}(x+y)&\ne f^{-1}_{\bxi}(x)\, f^{-1}_{\bxi}(y).
\end{align}
%
%
These two equations exhibit the underlying mathematical structure of
the generalized exponentials and their inverses, namely, the
generalized logarithms. Evidently, the introduction of the
deformation parameter $\bxi$ serves as a coupling between the
variables $x$ and $y$.

In this paper, we first show that deformed functions are
generally not invertible in the entire (co)domains of the ordinary functions
as a consequence of the aforementioned inequalities. In this sense,
the deformed functions used so far are found to be incomplete. Then,
we present the conditions under which bijectivity is satisfied.
Incorporating these conditions, we define complete generalized
logarithms and exponentials. Finally, we apply the criterion of
completeness to Tsallis, Kaniadakis, Abe and Borges-Roditi
generalized functions in order to elucidate some obscure points in
diverse mathematical constructions within the frame of generalized
BG-Statistics. As a result of our study, it becomes evident, that
the justification of whether a function can be considered as
\emph{generalized logarithmic} or \emph{generalized exponential}
function is associated  with some additional properties than just a
recovering limit $\bxi\ra\bxi_0$.

The paper is organized as follows. In Section \ref{sec:1} we present
the origin of the non-bijectivity of the $\bxi$-logarithm
(-exponential) from $\mathbb{R}^+$ ($\mathbb{R}$) to $\mathbb{R}$
($\mathbb{R}^+$). In order to assure the invertibility of the
$\bxi$-functions in the aforementioned sets, we introduce a
dependence between deformation parameters and specific subsets
$\mathbb{L}_i$ and $\mathbb{K}_i$ of $\mathbb{R}^+$
($\mathbb{R}^+=\bigcup_{i=1}^{n=2}\mathbb{L}_{i}$) and $\mathbb{R}$
($\mathbb{R}=\bigcup_{i=1}^{n=2}\mathbb{K}_{i}$), respectively.
Through the requirement of bijectivity, we are able to determine
explicitly the subsets $\mathbb{L}_i$ and $\mathbb{K}_i$. In
addition, we set forth conditions which permit us the determination
of the respective parameter validity ranges. In Section
\ref{applications}, we apply our formalism to Tsallis, Kaniadakis,
Abe and Borges-Roditi generalized functions in order to obtain
concomitant complete expressions. Finally, the main results will be
summarized in Section \ref{Concl}.

\section{Definition intervals of deformed functions}\label{sec:1}
%
We first define the following intervals
$\mathbb{L}_{0}:=\{x\in\mathbb{R}:0<x\leqslant1\}$,
$\mathbb{L}_{1}:=\{x\in\mathbb{R}:x\geqslant1\}$ so that
$\mathbb{L}_{0}\bigcup\mathbb{L}_{1}=$
$\mathbb{R}^+:=\{x\in\mathbb{R}:x>0\}$ i.e., the set of all positive
real numbers. We also choose the intervals $\mathbb{K}_1$ and
$\mathbb{K}_2$ as $\mathbb{R}^+_0:=\{x\in\mathbb{R}:x\geqslant0\}$
and $\mathbb{R}_0^-:=\{x\in\mathbb{R}:x\leqslant0\}$, respectively
so that $\mathbb{R}=\mathbb{R}^{-}_0\bigcup\mathbb{R}^{+}_0$ is the
set of all real numbers. We denote the parameter validity ranges of
the respective parameters $\bxi$ by
$\mc{A}_{\bxi}\subseteq\mathbb{R}$. A boundary value of
$\mc{A}_{\bxi}$ is always $\bxi_0$ e.g.,
$\mc{A}_{\bxi}:=(\alpha,\bxi_0]$ with $\alpha<\bxi_{0}$ or
$\mc{A}_{\bxi}:=[\bxi_0,\alpha)$ $\alpha>\bxi_{0}$. For the sake of
simplicity, we shall conduct our study starting from Eq.
\eqref{eq:1-logq}, although the same results can analogously be
obtained from Eq. \eqref{eq:2-expq}.

The ordinary logarithm (in natural base $e$)
$\ln:\mathbb{R}^+\ra\mathbb{R}$ has an argument intercept at the
point $x=1$, since $\ln(x=1)=0$, splitting the logarithmic domain
$\mathbb{R}^+$ into two subdomains $\mathbb{L}_0$ and
$\mathbb{L}_1$. The relation between the images of the log-arguments
$x\in\mathbb{L}_1$ and $1/x\in\mathbb{L}_0$ is represented by the
identity
%
\begin{align}
\label{ord-log}%
\ln(x)+\ln(1/x)&=0
\qquad\rbr{f_{\bxi_0}(x)+f_{\bxi_0}(1/x)=0}.
\end{align}
%
The right hand side of Eq. \eqref{ord-log}, being equal to zero,
guarantees that for $x\in\mb{R}^+$, the logarithmic codomain is
$\mb{R}$. This can be easily verified if we assume that the sum of the
above logarithmic terms does not cancel the $x$-dependence i.e.,
$\ln(x)=-\ln(1/x)+\Theta(x)$, where $\Theta$ is a real non-singular
function. Then, the codomain of the logarithmic function depends on
the codomain of $\Theta(x)$ as well, thus
$\ln:\mb{R}^+\ra\mb{M}\subseteq\mb{R}$ i.e., $\mb{M}$ denotes a subset of real numbers.

We now require the same (co)domain for $f_{\bxi}$ as that of
$f_{\bxi_0}$ and an analogous behavior of its image, namely,
$f_{\bxi}:\mathbb{L}_0\bigcup\mathbb{L}_1\ra\mathbb{R}_0^-\bigcup\mathbb{R}_0^+$,
$f_{\bxi}(x\in\mathbb{L}_0)\in\mathbb{R}_0^-$ and
$f_{\bxi}(x\in\mathbb{L}_1)\in\mathbb{R}_0^+$. Setting $y\ra1/x$ in
Eq. \eqref{eq:1-logq}, we are able to explore the relation between
the $x$- and $1/x$-arguments of $f_{\bxi}$. Then, for
$\bxi\ne\bxi_0$, we observe that $f_{\bxi}(x)+f_{\bxi}(1/x)\ne
f_{\bxi}(1)$. Since the term $f_{\bxi}(1)$ is a constant with
respect to the $x$-variable, and may take any value in $\mathbb{R}$,
the latter relation implies the existence of an $x$-dependent function
on the right hand side unlike the ordinary case i.e.,

\begin{align}
\label{eq:4-logq}%
f_{\bxi}(x)+f_{\bxi}(1/x)=\Theta_{\bxi}(x).
\end{align}
%
$f_{\bxi}$ represents a possible generalization only in a subset of
the codomain of the ordinary logarithmic function. In other words,
it is incomplete compared to its ordinary counterpart. Therefore, we
call a deformed function which is characterized by Eq.
\eqref{eq:4-logq} with $\Theta_{\bxi}(x)\ne0$ an \emph{incomplete
generalized logarithm},
$f_{\bxi}:\mb{R}^+\ra\mb{M}_{\bxi}\subseteq\mb{R}$, with its inverse
being \emph{incomplete generalized exponential} function i.e.,
$f^{-1}_{\bxi}:\mb{M}_{\bxi}\subseteq\mb{R}\ra\mb{R}^+$. An
incomplete definition has three major setbacks: first, as mentioned
above, the codomain $\mb{M}_{\bxi}$ is a subset of $\mb{R}$. This is
not a desired property, since it may restrict the range of the
argument interval in the $\bxi$-exponential $f^{-1}_{\bxi}$. It is
always possible that a particular physical system demands the use of
an argument external to the particular subset $\mb{M}_{\bxi}$. If
this is the case, then the applicability of the statistics based on
incomplete deformed functions will be impossible in the presumed
interval as a result of the underlying mathematical structure rather
than the requirements of the physical system under consideration.
Second, the subset $\mb{M}_{\bxi}$ depends on the deformation
parameters, providing a further restriction in the domain of
$f^{-1}_{\bxi}$. In other words, the applicability of the
generalized statistics is limited to a certain range depending on
the deformation parameter right from the beginning independent of
the physical system of interest. Therefore, it is ambiguous whether
the resulting deformation parameter range is a result of the
physical model or just chosen from the accessible range due to the
nature of the deformation parameter. Third setback is that the
function $f^{-1}_{\bxi}(x)$ does not describe a $\bxi$-exponential
decay for $x<0$ due to the dependence of $\mb{M}_{\bxi}$ on the
deformation parameters. This can be observed by inverting Eq.
\eqref{eq:4-logq}, which yields the relation $f_{\bxi
}^{-1}(-x)=1/f_{\bxi }^{-1}(x+\Theta _{\bxi }^{'} (x))$, where
$\Theta _{\bxi }^{'} (x):=\Theta _{\bxi }(1/f_{\bxi }^{-1}(-x))$.
This relation implies that $f^{-1}_{\bxi}(x)$ represents a
$\bxi$-exponential decay only when $\Theta _{\bxi }^{'} (x) = 0$.
The aforementioned observation can also be made by recourse to Eq.
\eqref{eq:2-expq}, since it yields the relation $f_{\bxi
}^{-1}(-x)f_{\bxi }^{-1}(x)\neq 1$ if we set $y\rightarrow-x$.

An immediate question then is whether one can define a complete
generalized logarithm. This consists of two steps: i) to eliminate
the $\bxi$-dependence of the codomain, $\mb{M}_{\bxi}\ra\mb{M}$, and
ii) to increase the range of the codomain, such that
$\mb{M}=\mb{R}$. Since the origin of the incompleteness lies in the
existence of nonzero $\Theta_{\bxi}(x)$ in Eq. \eqref{eq:4-logq},
the elimination of the latter function would satisfy the
aforementioned steps simultaneously. In order to do this, we assume
the existence of the functions
$d_k:\mc{A}_{\bxi}\ra\mc{B}^{(k)}_{\bxi}$
($\mc{B}^{(k)}_{\bxi}\subseteq\mathbb{R}$) with
$k=1,\ldots,v\geqslant u$ such that

\begin{align}\label{NewRel1a}
f_{\bxi}(x)+f_{d_{k}(\bxi)}(1/x)&=f_{\bxi}(1)
\end{align}
%
with $\lim_{\bxi\ra\bxi_0}d_{k}(\bxi)=\bxi_0$. The index $k$
associates a $d$-function to each parameter $\xi_j$, $v=u$. However,
it is possible to express the $i$th deformation parameter as a
function of the $\ell$th one ($j=1,\ldots,i,\ldots,\ell,\ldots,u$),
$\xi_i=\xi_i(\xi_{\ell})$, so that we have two different deformation
parametric structures based on the same deformation parameter. This
degeneracy is captured when $v>u$. Setting $x=1$ in Eq.
\eqref{NewRel1a}, we obtain $f_{d_k(\bxi)}(1)=0$ in accordance with
the ordinary logarithm. It can be seen that the above relation is
valid for any set of parameters by substituting $\bxi'=d_k(\bxi)$
into the above equation. Therefore, Eq. \eqref{NewRel1a} can be
rewritten as

\begin{align}\label{x-corresp}
f_{\bxi}(x)&=-f_{d_k(\bxi)}(1/x)
\end{align}
%
in analogy to Eq. \eqref{ord-log}. It is worth remarking that one can
obtain Eq. \eqref{x-corresp}, by changing the term $f_{\bxi}(1)$ in
Eq. \eqref{NewRel1a} to $f_{d_k(\bxi)}(1)$, as well. Because of Eq.
\eqref{eq:4-logq}, we have $d_k(\bxi)\ne\bxi$, which means that even
if the domain $\mc{A}_{\bxi}$ is equal to the codomain
$\mc{B}^{(k)}_{\bxi}$, the image of $d_k(\bxi)$ (except at the
boundary value $\bxi_0$) is not equal to the image of $\bxi$. We
denote $d$ as \emph{dual function} and the correspondence
$\bxi\leftrightarrow d_k(\bxi)$ as \emph{duality relation}. The
explicit structure of the dual mapping depends on the particular
expression of the generalized functions.

It should also be noted that the transformation of the arguments
i.e., $x\ra1/x$ given by Eq. \eqref{x-corresp}, keeping the
deformation parameters invariant, can equivalently be considered as
the transformation of the parameters as well, $\bxi\ra\tilde{\bxi}$
and $d_k(\bxi)\ra\tilde{d}_k(\tilde{\bxi})$, yielding the expression

\begin{align}\label{NewRel1b}
f_{\tilde{d}_k(\tilde{\bxi})}(x)&=-f_{\tilde{\bxi}}(1/x)
\end{align}
%
with $\bxi\ne\tilde{\bxi}$. The function $\tilde{d}_k$ has the same
expression with $d_k$, since the dual function is unique for each
$\bxi$-function, but presents the reverse (co)domain,
$\tilde{d}_k:\mc{B}^{(k)}_{\tilde{\bxi}}\ra\mc{A}_{\tilde{\bxi}}$, with
$\mc{A}_{\bxi}=\mc{A}_{\tilde{\bxi}}$ and
$\mc{B}^{(k)}_{\bxi}=\mc{B}^{(k)}_{\tilde{\bxi}}$. We further observe that the
argument $x=1$, since $f_{\bxi}(1)=0$, divides the (co)domain
$\mathbb{R}^+$ ($\mathbb{R}$) into the following sub(co)domains
$\mathbb{R}^+=\mathbb{L}_0\bigcup\mathbb{L}_1$
($\mathbb{R}=\mathbb{R}^-\bigcup\mathbb{R}_0^+$). Consequently, we
are led to the complete $\bxi$-logarithmic definition, $\ln_{\bxi}$, from
$\mathbb{R}^+$ to $\mathbb{R}$ with $\bxi\in\mc{A}_{\bxi}$ when the subsets of the (co)domain
correspond to different deformation parameters i.e.,
%
\begin{align}\label{(co)domain2-a}%
\ln_{\bxi}:=\begin{cases}
f_{d_k(\bxi)}:\mathbb{L}_0\longrightarrow\mathbb{R}^{-}_0\\
\\
f_{\bxi}:\mathbb{L}_1\longrightarrow\mathbb{R}^{+}_0
\end{cases}
\qquad\mt{with}\qquad
\exp_{\bxi}\equiv\ln^{-1}_{\bxi}:=
\begin{cases}
f^{-1}_{d_k(\bxi)}:\mathbb{R}^{-}_0\longrightarrow\mathbb{L}_0\\
\\
f^{-1}_{\bxi}:\mathbb{R}^{+}_0\longrightarrow\mathbb{L}_1
\end{cases}\,.
\end{align}
%
We can then, for the complete generalized logarithm and exponential
given above, verify that
%
\begin{align*}
\ln_{\bxi}(1/x)&=f_{d_k(\bxi)}(1/x)=-f_{\bxi}(x)=-\ln_{\bxi}(x),\\
\exp_{\bxi}(-x)&=f^{-1}_{d_k(\bxi)}(-x)=\frac{1}{f^{-1}_{\bxi}(x)}=\frac{1}{\exp_{\bxi}(x)}
\end{align*}
%
for $x\in\mb{L}_1$ and $x\in\mb{R}^+_0$, respectively. On the other
hand, when a deformed logarithm fulfills the relation in Eq.
\eqref{eq:4-logq} with $\Theta_{\bxi}(x)=0$, the transformation
$x\ra1/x$ does not cause any structural changes, implying that
$f_{\bxi}$ is $\bxi$-symmetric (i.e., $f_{\bxi}$ and $f_{d_k(\bxi)}$
have the same images). If this is the case, the deformed functions
can be defined analogous to the ordinary logarithmic and exponential
functions (see Section \ref{sec:3} for such a case). Combining Eqs.
\eqref{eq:4-logq} and \eqref{x-corresp}, we are able to construct a
criterion based on the dual function $d_k(\bxi)$ and the primary
deformed function $f_{\bxi}$. A deformed function $f_{\bxi}$
represents a single-parameter-set complete and analytic generalized logarithm,
$f_{\bxi}:\mathbb{R}^+\ra\mathbb{R}$ ($f_{\bxi}\equiv\ln_{\bxi}$),
when the following relation is satisfied
%
\begin{align}\label{cond-dual}
f_{d_k(\bxi)}(x)&=f_{\bxi}(x)\qquad \mt{for}\qquad d_k(\bxi)\ne\bxi.
\end{align}
%
The same criterion can be expressed for the inverse deformed
function $f^{-1}_{\bxi}$ through the relation
$f^{-1}_{\bxi}(-x)f^{-1}_{d_k(\bxi)}(x)=1$.

The parameter range $\mc{A}_{\bxi}$ can be
determined by requiring the fulfillment of the following limits
satisfied by the ordinary functions
%
\begin{subequations}\label{sec2-eq:C}
\begin{align}
\label{GenCon1}%
\lim_{x\ra0}\exp_{\bxi}(x)&=1, &\lim_{x\ra1}\ln_{\bxi}(x)&=0,\\
\label{GenCon2}%
\lim_{x\ra-\infty}\exp_{\bxi}(x)&=0,&\lim_{x\ra0}\ln_{\bxi}(x)&=-\infty,\\
\label{GenCon3}%
\lim_{x\ra\infty}\exp_{\bxi}(x)&=\infty,&\lim_{x\ra\infty}\ln_{\bxi}(x)&=\infty,\\
\label{GenCon4}%
\lim_{x\ra-\infty}\frac{d}{dx}\exp_{\bxi}(x)&=0,&\lim_{x\ra\infty}\frac{d}{dx}\ln_{\bxi}(x)&=0.
\end{align}
\end{subequations}

Condition \eqref{GenCon1} implies that $\ln_{\bxi}(x)$ and
$\exp_{\bxi}(x)$ are continuous functions at the points $x=1$ and
$x=0$, respectively. Eqs. \eqref{GenCon2} and \eqref{GenCon3}
describe the behavior of the generalized functions at the boundaries
of their domains. The condition \eqref{GenCon4} ensures the
preservation of the same absolute maximum and minimum of the
ordinary functions. As we shall see in the next section, the
conditions listed in Eq. \eqref{sec2-eq:C} are in complete agreement
with the definitions in Eq. \eqref{(co)domain2-a}, yielding a
complete definition of the generalized functions.

\section{Applications}\label{applications}

Having introduced the criterion of the completeness in the context
of the generalized functions, we now apply it to the Tsallis,
Kaniadakis, Abe and Borges-Roditi generalized functions in order to
obtain concomitant complete expressions in case that they are not
already complete. These applications will not only show the
applicability of the criterion developed here but also indicate the
general nature of the current formalism. Moreover, the criterion
of completeness naturally yields the concavity interval of the
R\'enyi entropy if one defines the R\'enyi entropy through
$q$-deformed generalized functions and thereby using the
completeness of these functions.

\subsection{Tsallis $q$--functions}\label{sec:2}

A one parameter ($\bxi=\{\xi_1\}=: q$ with $q_0=1$) generalization
introduced by Tsallis and coworkers \cite{Tsallis1988,Tsnewbook} reads
%
\begin{equation}\label{Ts-functions}
f_{q}(x):=\frac{x^{1-q}-1}{1-q},\qquad\qquad h_{q}(x):=\Big[1+(1-q)\,x\Big]^{\frac{1}{1-q}},
\end{equation}

with $f_{q\ra1}(x)=\ln(x)$ and $h_{q\ra1}(x)=\exp(x)$. It can be
easily shown that $f_q(x)\ne-f_q(1/x)$, since $\Theta_q(x)$
$=\frac{x^{1-q}+x^{q-1}-2}{1-q}\neq0$ (or equivalently
$h_q(x)h_q(-x)\ne1$ where $h_q(x)\equiv f^{-1}_q (x)$). Concerning
their domain and codomain, we observe
%
\begin{subequations}\label{Ts-fun-(co)do}
\begin{align}
q<1& \qquad f_q:\mathbb{R}_0^+\longrightarrow \bigg[\frac{1}{q-1},\infty\bigg),
\qquad\quad\,%
h_q:\bigg[\frac{1}{q-1},\infty\bigg)\longrightarrow\mathbb{R}^+_0,\\
q=1& \qquad f_q:\mathbb{R}^+\longrightarrow \mathbb{R},
\qquad\quad\quad\;\,\,\quad\qquad%
h_q:\mathbb{R}\longrightarrow\mathbb{R}^+,\\
q>1& \qquad f_q:\mathbb{R}^+\longrightarrow \bigg(-\infty,\frac{1}{q-1}\bigg],
\qquad%
h_q:\bigg(-\infty,\frac{1}{q-1}\bigg]\longrightarrow\mathbb{R}^+.
\end{align}
\end{subequations}
%

The generalized functions, $f_q$ and $h_q$, are not complete when
$q\ne1$ as can be seen from Eq. \eqref{Ts-fun-(co)do}. This limits
the use of deformed functions given by \eqref{Ts-functions} to some
subsets of the real numbers for ($q\ne1$), not as a result of the
physical system under investigation but solely on the ground of the
incompleteness of the underlying mathematical structure. Since one
first needs the dual function $d_{k=1}(q)=d(q)$ in order to define complete
generalized functions, we calculate it from Eqs. \eqref{x-corresp}
and \eqref{Ts-functions} as

\begin{align}\label{Ts-DualRel}
d(q)&=2-q.
\end{align}
%
We notice that the analytical expression in Eq. \eqref{Ts-DualRel}
depends as much on $q$ as on the recovering value $q_0$. For
example, the reparametrization $q'=1-q$ in the above equation
changes the dual mapping $d(q)$ to $d'(q')=-q'$ with $q'_0=0$.
The requirements listed in Eq. \eqref{sec2-eq:C} confines the values
of the deformation parameter $q$ into the following interval

\begin{equation}\label{q-interval-a}
\mc{A}_q:=(0,1].
\end{equation}
%
Note that $\mc{A}_q$ coincides with the parameter interval obtained from the normalization of the q-exponential decay i.e., $\int_0^\infty dx/h_q(x){=}1$.

Having explicitly obtained the dual mapping function $d(q)$ and the
range of validity of the deformation parameter $q$, we can now
define the complete $q$-generalized functions
%
\begin{align}\label{Ts-fun-2}%
\ln_{q}(x):=\begin{cases}
\displaystyle\frac{x^{q-1}-1}{q-1}, & x\in\mb{L}_0\\
\\
\displaystyle\frac{x^{1-q}-1}{1-q}, & x\in\mb{L}_1
\end{cases}\,,
\qquad\qquad
\exp_{q}(x):=
\begin{cases}
\Big[1+(q-1)\,x\Big]^{\frac{1}{q-1}}, & x\in\mb{R}_0^-\\
\\
\Big[1+(1-q)\,x\Big]^{\frac{1}{1-q}}, & x\in\mb{R}_0^+
\end{cases}\,,
\end{align}
%
%
in accordance with Eq. \eqref{(co)domain2-a}. These expressions are also obtained, with a slightly different choice, by Teweldeberhan \textit{et al.} in Ref.\cite{Teweldeberhan} in the context of cut-off prescriptions associated with the $q$-generalized functions. These authors too have noticed that a division of the (co)domains yields to a complete definition of the $q$-exponentials. However, this observation has not been pursued further as a criterion of completeness in Ref. \cite{Teweldeberhan}.

\begin{figure}[b]
\begin{center}
    \includegraphics[width=12cm,height=10cm]{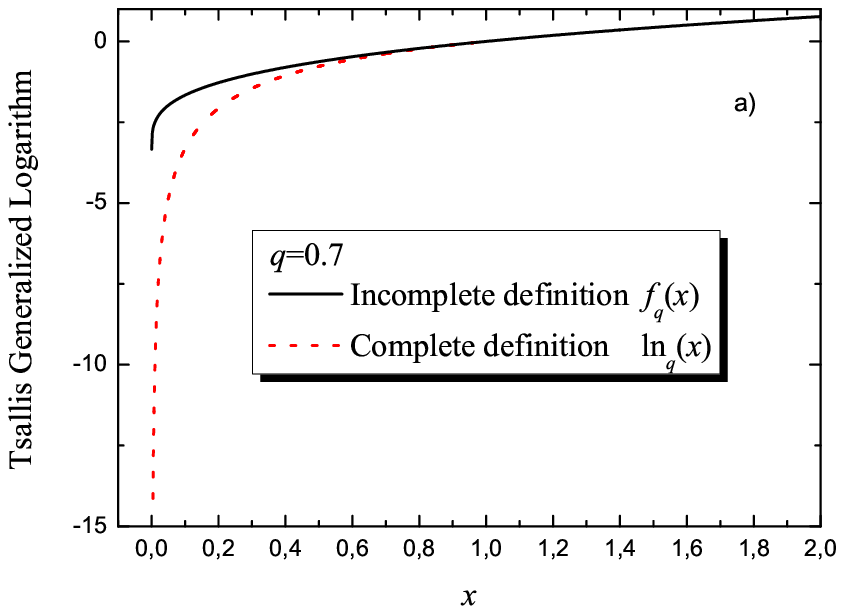}
    \includegraphics[width=12cm,height=10cm]{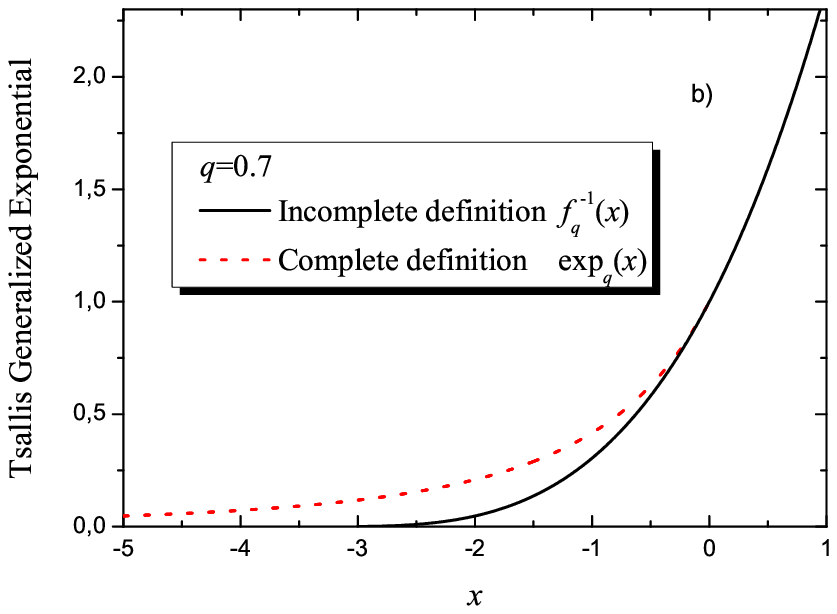}
        \caption{a) Plot of Tsallis incomplete ($f_q$, solid line) and complete ($\ln_q$, dotted line) deformed logarithm for $q=0.7$.
             b) Plot of Tsallis incomplete ($f^{-1}_q$, solid line) and complete ($\exp_q$, dotted line) deformed exponential for $q=0.7$.}%
  \label{Tsallis.Log}
\end{center}
\end{figure}

We plot the incomplete (solid lines) and complete (dashed lines)
$q$-generalized logarithms in Eqs. \eqref{Ts-functions} and
\eqref{Ts-fun-2}, respectively, corresponding to the deformation
parameter $q=0.7$, in Fig. \ref{Tsallis.Log}a. The incomplete
$q$-logarithm $f_q$ has a cut-off at the point $x=0$ corresponding
to the value $\frac{1}{q-1}$, which has no analog in the ordinary
logarithmic function. The complete $q$-logarithm $\ln_q$, on the
other hand, behaves as the ordinary logarithmic function, tending to
$\ln_{q}(x)\ra-\infty$ as $x\rightarrow0$. The complete and
incomplete $q$-logarithms exhibit the same behavior for $x\geq1$ as
expected, since they are defined from the same domain to the same
codomain. For the interval $x<1$, the incomplete generalized
logarithm reaches the cutoff and stabilizes therein, whereas the
complete $q$-logarithm, mapped by the dual function, smoothly
approaches to $-\infty$. An inspection of Fig. \ref{Tsallis.Log}b
reveals the analogous results for the incomplete and complete
generalized exponentials for $q=0.7$.

\subsubsection{R\'enyi entropy}
One of the most discussed deformed entropic structures introduced
first within information theory and then as a possible
generalization of Boltzmann-Gibbs thermostatistics is the one
defined by R\'enyi \cite{Renyi1970,LenziMendesSilva}

\begin{equation}\label{R-ent}
S_r=\frac{1}{1-r}\ln\rbr{\sum_{i}p_i^r}
\end{equation}
%
where $r$ is the deformation parameter and $S_{r\ra1}=\BG$. It has
been previously shown that it exhibits a well-defined concavity only
for $r\in(0,1)$ \cite{R-Concavity}. The nonconcavity of the R\'enyi
entropy outside the aforementioned range has been frequently used as
an argument for the inappropriateness of the $S_r$-definition as a
candidate for BG-generalization.

BG-entropy can be derived, without appealing to physical laws, by
considering the asymptotic behavior of the Ordinary Multinomial
Coefficients (OMC). The OMC gives the number of all possible
configurations -- or (micro)states in physics -- created from the
combination of objects without repetitions. In a recent study
\cite{Oikonomou07}, one of the present authors (TO) constructed
properly defined Deformed Multinomial Coefficients (DMC), based on
the concept of the deformation parameters, in order to derive
Tsallis, R\'enyi and nonextensive Gaussian \cite{Oik2006a}
entropies. Demanding the positivity of DMC, since a negative number
of configuration is senseless, it has been demonstrated that R\'enyi
entropy is defined only for $r\in(0,1)$. The same result about the
parameter validity range is obtained in Ref. \cite{BagciTirnakli} by
applying Jaynes' formalism for $S_r$ with ordinary (in contrast to
the escort) exponentially averaged constraints in accordance with
the fundamental mathematical structure of the R\'enyi entropy.

We now want to derive the interval of concavity in a more
fundamental way following a different path, related to the parameter
range in Eq. \eqref{q-interval-a} and the definitions given by Eq.
\eqref{Ts-fun-2}. In Ref. \cite{Oikonomou07}, it has been
established that R\'enyi entropy can be expressed through the
deformed multiplication introduced in Ref. \cite{Borges04} within
Tsallis statistics [$x\otimes_r y:=\exp_r(\ln_r(x)+\ln_r(y))$,
$x,y>1$] as follows
%
\begin{equation}\label{R-ent}
S_r=\ln\rbr{\tprod{r}_i\rbr{1/p_i}^{\otimes_{r}^{p_i}}},\qquad
\tprod{r}_{i}\rbr{1/p_i}^{\otimes_{r}^{p_i}}=\exp_r\rbr{\sum_ip_i\ln_r(1/p_i)}=\sqbr{\sum_ip_i^r}^{\frac{1}{1-r}}.
\end{equation}
%

Assuming that the deformed functions in Eq. \eqref{R-ent} are
complete, we see that the $\ln_r$-argument takes values in
$\mathbb{L}_1$ (i.e., the microprobabilities vary between zero and
one), and thus the $\exp_r$-argument varies in the interval
$\mathbb{R}_0^+$. It becomes then evident that the deformation
parameter $r$ takes values in $(0,1)$ as a consequence of the
mathematical definition of the generalized functions in Eq.
\eqref{Ts-fun-2}. Accordingly, R\'enyi entropy is concave for
all values of its parameter range, when the complete deformed
functions are being used instead of the incomplete ones. However,
the very fact that one uses the complete generalized functions
restricts the parameter range only to values in the interval
$(0,1)$. Once we use the complete deformed functions rather than the
incomplete ones, the interval of the concavity coincides with the
interval of the parameter range.

\subsection{Kaniadakis $\kappa$--functions}\label{sec:3}
\begin{figure}[b]
\begin{center}
  \includegraphics[width=12cm,height=10cm]{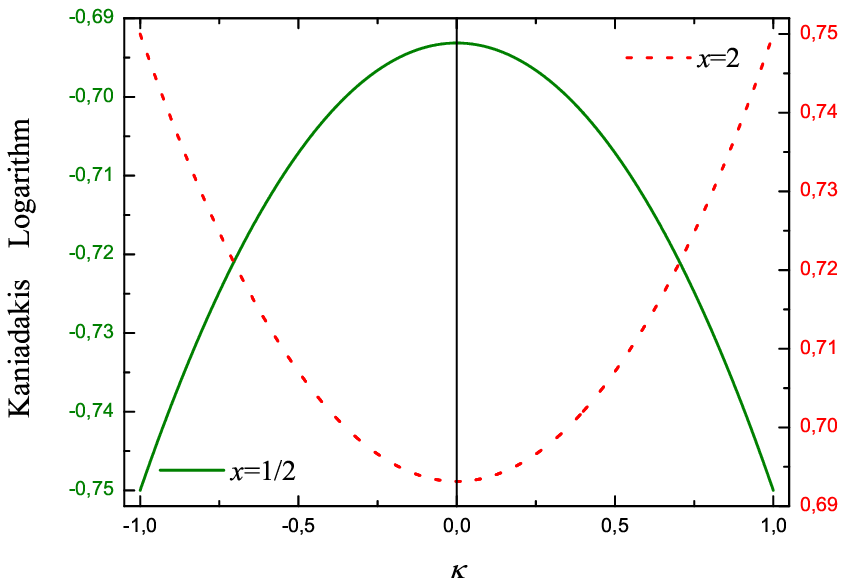}
     \caption{Plot of Kaniadakis logarithm with regard to the deformation parameter $\kappa$ in the interval
            $\mc{A}_\kappa\bigcup\mc{B}_\kappa=(-1,0]\bigcup[0,1)$ for $x=2$ (dotted, right axis) and $x=1/2$ (solid, left axis).}%
  \label{Kaniadakis.Log}
\end{center}
\end{figure}

%
%
Mathematically, a very interesting case is
represented by the one parametric ($\bxi=\{\xi_1\}=:\kappa$ with $\kappa_0=0$) generalization
introduced by Kaniadakis \cite{Kaniadakis02,KaniadQuaratiScarf,Kaniadakis05}. It is based on the following deformed functions

\begin{align}\label{K-functions}
f_{\kappa}(x):=\frac{x^{\kappa}-x^{-\kappa}}{2\kappa},\qquad
h_{\kappa}(x):=\Big[\sqrt{1+\kappa^2x^2}+\kappa x\Big]^{1/\kappa}
\end{align}
%
with $f_\kappa:\mathbb{R}^+\lra\mathbb{R}$ and
$h_\kappa:\mathbb{R}\lra\mathbb{R}^+$ for $\forall\kappa\in\mathbb{R}$.
Then, since $f\circ h=\mt{id}_{\mathbb{R}}$ and $h\circ
f=\mt{id}_{\mathbb{R}^+}$, $h$ is the inverse of $f$ for all allowed
values of $x$ and $\kappa$ i.e., $h_\kappa\equiv f^{-1}_\kappa$. The
peculiarity of the $\kappa$--functions is that they satisfy
simultaneously Eqs. \eqref{eq:1-logq} and \eqref{eq:2-expq} as well
as the relations
%
\begin{equation}\label{K-Peculiarity}
f_{\kappa}(x)=-f_{\kappa}(1/x)\qquad\Leftrightarrow\qquad f^{-1}_{\kappa}(x)f^{-1}_{\kappa}(-x)=1.
\end{equation}
%
It can be easily verified that a transformation $x\ra1/x$ of the
$\kappa$-logarithmic argument or a transformation $x\ra-x$ of the
$\kappa$-exponential argument in Eq. \eqref{K-Peculiarity} holds the
respective relation unaltered, $\kappa=\tilde{\kappa}$ and
$d_{k=1}(\kappa)=\tilde{d}_{k=1}(\tilde{\kappa})$. This means that
the Kaniadakis functions are $\kappa$-symmetric and thus they
preserve the same parameter (or parameter range) in their domains
as well as in their codomains. The definitions in Eq. \eqref{K-functions} are the
only ones known up to now in literature which present the
aforementioned property. The same result is obtained using the
formalism in Section \ref{sec:1}. Assuming the expressions in Eq.
\eqref{K-Peculiarity} are not valid, we can still look for a function
$d_{k=1}\equiv d:\mc{A}_\kappa\lra\mc{B}_\kappa\equiv\mc{B}^{(1)}_{\kappa}$ such that
%
\begin{equation}\label{K-form}
f_{\kappa}(x)=-f_{d(\kappa)}(1/x)\qquad\Leftrightarrow\qquad f^{-1}_{\kappa}(x)f^{-1}_{d(\kappa)}(-x)=1.
\end{equation}
%
From Eqs. \eqref{K-functions} and \eqref{K-form}, we determine the form of $d$ as
%
\begin{align}\label{K-DualRel}
d(\kappa)&=-\kappa.
\end{align}
%
Comparing the $\kappa$- and $d(\kappa)$-structure of $f$ and $f^{-1}$ we obtain (see the criterion of completeness i.e., Eq. \eqref{cond-dual})
%
\begin{equation}\label{k-symmetry}
f_{d(\kappa)}(*)=f_{\kappa}(*)\qquad\Leftrightarrow\qquad f^{-1}_{d(\kappa)}(*)=f^{-1}_{\kappa}(*),
\end{equation}
%
where * represents any argument of $f$ or $f^{-1}$. Accordingly, the
assumption does not hold and the relations in Eq.
\eqref{K-Peculiarity} are indeed satisfied. From Eq. \eqref{sec2-eq:C} and
the relation $d(\mc{A}_\kappa)=\mc{B}_\kappa$ we are able to
determine the $\kappa$- and $d(\kappa)$-interval given as follows
%
\begin{equation}\label{kappa-interval-b}
\mc{A}_\kappa:=[0,1),\qquad \mc{B}_\kappa:=(-1,0].
\end{equation}
%
Eq. \eqref{k-symmetry} implies that the functions in Eq.
\eqref{K-functions} are $\kappa$--symmetric, they exhibit the same
image for $\kappa$ and $d(\kappa)$. Thus, whether we make use of
interval $\mc{A}_\kappa$ or of interval $\mc{B}_\kappa$, we obtain
the same result. In Fig. \ref{Kaniadakis.Log}, we demonstrate this
behavior letting the deformation parameter $\kappa$ vary in the
united range
$(-1,1)=\mc{B}_\kappa\bigcup\mc{A}_\kappa=d(\mc{A}_\kappa)\bigcup\mc{A}_\kappa$
for $x=2$ and $1/x=2$. We note that the intervals $\mc{A}_\kappa$
and $\mc{B}_\kappa$ in Eq. \eqref{kappa-interval-b} are identical
with the ones obtained from the normalization of
$1/f^{-1}_{\kappa}(x)$ and $f^{-1}_{\kappa}(-x)$, respectively.

As a result of the above discussion, we accomplish the definitions in Eq. \eqref{K-functions} with the following information
%
\begin{subequations}
\begin{align}
f_\kappa\equiv\ln_\kappa&:\mathbb{R}^+\lra\mathbb{R}\qquad\mt{with}\qquad \kappa\in\mc{A}_\kappa\vee\kappa\in\mc{B}_\kappa,\\
f^{-1}_\kappa\equiv\exp_\kappa&:\mathbb{R}\lra\mathbb{R}^+\qquad\mt{with}\qquad \kappa\in\mc{A}_\kappa\vee\kappa\in\mc{B}_\kappa.
\end{align}
\end{subequations}
%
As can be seen, the domain and the codomain of $f_\kappa$ and
$f^{-1}_\kappa$ are completely analogous to the undeformed ones
i.e., $\ln$ and $\exp$, respectively. In Ref. \cite{Kaniadakis02},
Kaniadakis showed that these deformed functions exhibit similar
behaviors as the ordinary, undeformed functions for all boundary
values of $\mathbb{R}$ and $\mathbb{R}^+$.
%
%
%

\subsection{Abe $q$--logarithm}\label{sec:Abe}
%
Abe's one-parameter generalized logarithmic function
($\bxi=\{\xi_1\}=: q$ with $q_0=1$) \cite{Abe97} reads
%
\begin{equation}\label{Abe-functions}
f_{q}(x):=\frac{x^{1/q-1}-x^{q-1}}{1/q-q},\qquad\qquad
h_{q}(x):=f^{-1}_q(x)
\end{equation}
%
with $f_{q\ra1}(x)=\ln(x)$ and $h_{q\ra1}(x)=\exp(x)$. The
$q$-exponential function is not analytically invertible for all
allowed values of $q$. Similar to the Tsallis case, the current
definition is incomplete since $f_q(x)\ne-f_q(1/x)$ with
$\Theta_q(x)=\frac{x^{1/q-1}-x^{q-1}+x^{1-1/q}-x^{1-q}}{1/q-q}$.
Concerning its domain and codomain, we obtain
%
\begin{subequations}\label{Abe-fun-(co)do}
\begin{align}
q>0& \qquad f_q:\mathbb{R}^+\longrightarrow \mb{R},\\
q<0& \qquad f_q:\mathbb{R}^+\longrightarrow \mathbb{R}_0^{-},\\
q=0& \qquad f_q:(0,1]\bigcup(1,\infty)\longrightarrow 0\bigcup\infty.
\end{align}
\end{subequations}
%
We observe an implicit dependence of the codomains on the
deformation parameter in the equation above, in contrast to the
explicit parameter dependence in Tsallis case. The peculiarity of
Abe's definition is that there exists a parameter variation range
($q>0$), where the codomain is identified with the one of the
ordinary logarithm ($q=1$). However, one should be aware that this
property does not imply completeness of $f_q(x)$ for $q>0$, since $\Theta_q(x)\ne0$ (see the third setback in Section \ref{sec:1}).

The validity range of the deformation parameter is calculated
through the conditions in Eq. \eqref{sec2-eq:C}, which yields
%
\begin{align}
\mc{A}_q:=(1/2,1]\bigcup[1,2).
\end{align}
%
The duality relations are determined from Eqs. \eqref{x-corresp} and
\eqref{Abe-functions}
%
\begin{align}\label{Abe-DualRel}
d_1(q)&=\frac{q}{2q-1},\qquad d_2(q)=2-q.
\end{align}
%
The existence of two dual functions, $f_q(x)=-f_{d_1(q),d_2(q)}(1/x)$, depending on the same deformation parameter implies that Abe's logarithm issues from a structure which consists primarily of two different parameters (see next subsection).

Having explicitly obtained the dual mapping function $d(q)$ and the
range of validity of the deformation parameter $q$, we can now give
the definition of Abe's complete $q$-generalized functions
%
\begin{align}\label{Abe-fun-2}
\ln_{q}(x):=\begin{cases}
\displaystyle\frac{x^{1-1/q}-x^{1-q}}{q-1/q}, & x\in\mb{L}_0\\
\\
\displaystyle\frac{x^{1/q-1}-x^{q-1}}{1/q-q}, & x\in\mb{L}_1
\end{cases}\,,
\qquad\qquad
\exp_{q}(x):=
\begin{cases}
f^{-1}_{d_1(q),d_2(q)}(x), & x\in\mb{R}^-\\
\\
f^{-1}_q(x), & x\in\mb{R}_0^+
\end{cases}\,,
\end{align}
%
in accordance with Eq. \eqref{(co)domain2-a}.

\begin{figure}[b]
\begin{center}
  \includegraphics[width=12cm,height=10cm]{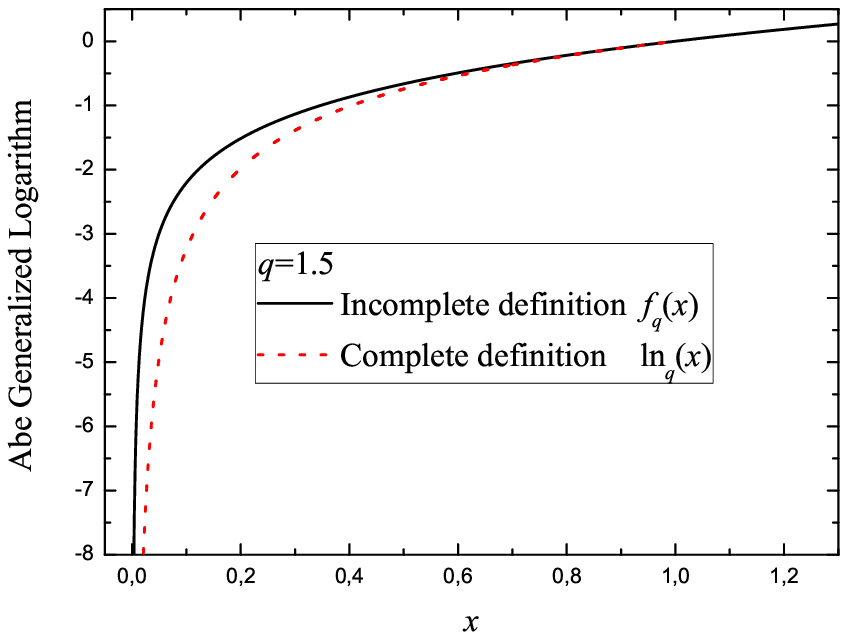}
    \caption{Plot of Abe's incomplete ($f_q$, solid line) and complete ($\ln_q$, dotted line) deformed
                logarithm, for $q=1.5$.}%
  \label{Abe.Log}
\end{center}
\end{figure}

In Fig. \ref{Abe.Log}, we plot the incomplete (solid lines) and
complete (dashed lines) $q$-generalized logarithms in Eqs.
\eqref{Abe-functions} and \eqref{Abe-fun-2} for $q=0.6$,
respectively. Both functions tend to minus infinity when $x\ra0$,
which is justified through their (co)domain. However, $\ln_q$ decays
smoother to $-\infty$ than $f_q$ for $0<x<1$. The complete and
incomplete $q$-logarithms exhibit the same behavior for $x\geq1$ as
expected, since they are defined as a mapping from the same domain
to the same codomain.

%

\subsection{Borges-Roditi two-parametric
logarithm}\label{sec:BorgesRoditi}
%
A two-parameter ($\bxi=\{\xi_1,\xi_2\}$ with
$\xi_{1_0}=0=\xi_{2_0}$) incomplete generalization introduced by
Borges and Roditi (BR) \cite{BorgesRoditi} reads
%
\begin{equation}\label{BR-functions}
f_{\xi_1,\xi_2}(x):=\frac{x^{\xi_1}-x^{\xi_2}}{\xi_1-\xi_2},\qquad\qquad
h_{\xi_1,\xi_2}(x):=f^{-1}_{\xi_1,\xi_2}(x)
\end{equation}
%
with $f_{\xi_{1_0},\xi_{2_0}}(x)=\ln(x)$, $h_{\xi_{1_0},\xi_{2_0}}(x)=\exp(x)$ and $\Theta_{\xi_1,\xi_2}(x)=\frac{x^{\xi_1}-x^{\xi_2}+x^{-\xi_1}-x^{\xi_2}}{\xi_1-\xi_2}$. The generalized exponential function $h_{\xi_1,\xi_2}$ is not analytically invertible for all values of the deformation parameters. Concerning the domain and codomain of $f_{\xi_1,\xi_2}$, we observe
%
\begin{subequations}\label{BR-fun-(co)do}
\begin{align}
\xi_1,\xi_2>0& \qquad f_{\xi_1,\xi_2}:\mathbb{R}_0^+\longrightarrow [0,\infty),
\qquad\qquad\;\:\,%
h_{\xi_1,\xi_2}:[0,\infty)\longrightarrow\mathbb{R}^+_0,\\
\xi_1,\xi_2<0& \qquad f_{\xi_1,\xi_2}:\mathbb{R}^+\longrightarrow (-\infty,0],
\qquad\quad\;\;\:\,%
h_{\xi_1,\xi_2}:(-\infty,0]\longrightarrow\mathbb{R}^+,\\
\label{eq:12c}%
\xi_1>0, \xi_2=0& \qquad f_{\xi_1,\xi_2}:\mathbb{R}^+_0\longrightarrow (-1/\xi_1,\infty],
\qquad\;\;\:%
h_{\xi_1,\xi_2}:(-1/\xi_1,\infty]\longrightarrow\mathbb{R}^+_0,\\
\label{eq:12d}%
\xi_1<0, \xi_2=0& \qquad f_{\xi_1,\xi_2}:\mathbb{R}^+\longrightarrow (-\infty,-1/\xi_1],
\qquad%
h_{\xi_1,\xi_2}:(-\infty,-1/\xi_1]\longrightarrow\mathbb{R}^+,\\
\substack{\displaystyle\mt{sgn}(\xi_1)\ne\mt{sgn}(\xi_2)\\ \displaystyle\xi_1=0=\xi_2}& \qquad f_{\xi_1,\xi_2}:\mathbb{R}^+\longrightarrow \mb{R},
\qquad\qquad\qquad\;\:\,%
h_{\xi_1,\xi_2}:\mb{R}\longrightarrow\mathbb{R}^+.
\end{align}
\end{subequations}
%
The results in Eqs. \eqref{eq:12c} and \eqref{eq:12d} are analogous
for $\xi_1=0,\xi_2\gtrless0$. In Eq. \eqref{BR-fun-(co)do} it
becomes evident that the complete definition, in the sense of
Kanadiakis definition, of the functions $f_{\xi_1,\xi_2}$ and
$h_{\xi_1,\xi_2}$ may be found  only when
$\mt{sgn}(\xi_1)\ne\mt{sgn}(\xi_2)$ ($\xi_1=0=\xi_2$ corresponds to
the ordinary functions). We calculate the dual functions from Eqs. \eqref{x-corresp} and \eqref{BR-functions} as

\begin{align}\label{BR-DualRel}
d_1(\xi_1)&=-\xi_1, \qquad d_2(\xi_2)=-\xi_2.
\end{align}
%

The conditions in Eq. \eqref{sec2-eq:C} confines the values
of the deformation parameters $\xi_1$ and $\xi_2$ into the following intervals

\begin{subequations}\label{BRq-interval-a}
\begin{align}
\mc{A}_{\xi_{\alpha}}&:=(-\infty,0)\qquad\Leftrightarrow\qquad\mc{A}_{\xi_{\beta}}:=(0,1),\\
\mc{A}_{\xi_{\alpha}}&:=(0,1)\qquad\quad\,\Leftrightarrow\qquad\mc{A}_{\xi_{\beta}}:=(0,1),\\
\mc{A}_{\xi_{\alpha}}&:=[0,0]\qquad\quad\;\Leftrightarrow\qquad\mc{A}_{\xi_{\beta}}:=[0,1),
\end{align}
\end{subequations}
%
where $\alpha,\beta=1,2$ with $\alpha\ne\beta$. These results are in
agreement with the ones obtained in Eq. \eqref{BR-fun-(co)do} for
the domain of the argument when ($\xi_1,\xi_2\gtrless0$). The
boundary value, $1$, is obtained from the condition \eqref{GenCon4}.

Having explicitly obtained the dual mapping function $d(q)$ and the
range of validity of the deformation parameter $q$, we can now
define the complete $q$-generalized functions
%
\begin{align}\label{BR-fun-2}
\ln_{\xi_1,\xi_2}(x):=\begin{cases}
\displaystyle\frac{x^{-\xi_1}-x^{-\xi_2}}{\xi_2-\xi_1}, & x\in\mb{L}_0\\
\\
\displaystyle\frac{x^{\xi_1}-x^{\xi_2}}{\xi_1-\xi_2}, & x\in\mb{L}_1
\end{cases}\,,
\qquad\qquad
\exp_{\xi_1,\xi_2}(x):=
\begin{cases}
f^{-1}_{d_1(\xi_1),d_2(\xi_2)}(x), & x\in\mb{R}^-\\
\\
f^{-1}_{\xi_1,\xi_2}(x), & x\in\mb{R}_0^+
\end{cases}\,,
\end{align}
%
in accordance with Eq. \eqref{(co)domain2-a}.

A special feature of BR-logarithmic structure is that it includes
all three deformed logarithmic definitions we have studied in this
section. It can be verified that for $\{\xi_1=1-q,\,\xi_2=0\}$,
$\{\xi_1=\kappa=-\xi_2\}$ and $\{\xi_1=1/q-1,\,\xi_2=q-1\}$, one
obtains Tsallis, Kaniadakis and Abe logarithms, respectively.
Indeed, the properties of BR-function reproduce the ones of the
aforementioned functions. Thus, the BR generalized logarithms can be
characterized as a family of deformed logarithms. Considering the
$\Theta_{\xi_1,\xi_2}$-function, we observe that the BR-logarithmic
family includes only one possible structure which is complete and
analytic in all points of its domain, namely, when $\xi_2=-\xi_1$,
which implies $\Theta_{\xi_1,\xi_2}(x)=0$, as expected. This
corresponds to the Kanadiakis definition.

\section{Conclusions}\label{Concl}
%
The recent generalizations of Boltzmann-Gibbs Statistics are based
on the introduction of some deformed forms of the ordinary
logarithmic and exponential functions, where
$\bxi=\{\xi_i\}_{i=1,\ldots,u}$ denotes the deformation parameter
set. It is hence of great importance to understand the mathematical
structure of these deformed functions in order to gain more insight
into the associated generalization schemes. Motivated by this fact,
we first demonstrated that the generalized logarithms $f_{\bxi}$
(and their inverse i.e., generalized exponentials $h_{\bxi}$) are
non-bijective from $\mathbb{R}^+$ ($\mathbb{R}$) to $\mathbb{R}$
($\mathbb{R}^+$). Thus, the generalized logarithm definitions
$f_{\bxi}$ may represent the inversion of the generalized
exponentials $h_{\bxi}$ and vice versa, only in some subsets of
$\mb{R}^+$ and $\mb{R}$. Therefore, they are called incomplete
deformed functions. This feature issues from the non-additive
$\bxi$-logarithmic and non-multiplicative $\bxi$-exponential
composition rules i.e., $f_{\bxi}(xy)\ne f_{\bxi}(x)+f_{\bxi}(y)$
and $h_{\bxi}(x+y)\ne h_{\bxi}(x)h_{\bxi}(y)$. The incompleteness of
deformed functions is not only a mathematical deficiency, but also
presents some physical problems, since it restricts the
applicability of the associated thermostatistics. In other words, it
is the incomplete mathematical structure, which dictates the range
of arguments or results, rather than the physical system one
studies. However, the estimation of the aforementioned subsets is
not a trivial procedure, since some boundary values present
$x(\bxi)$-dependence. In order to overcome this difficulty and to
guarantee the property of bijectivity in the original (co)domains,
we proposed dual mapping functions $d_k(\bxi)$ ($k=1,\ldots,v\geqslant u$), which enable a
change of the parameters within specific subsets of $\mb{R}^+$ and
$\mb{R}$. Through the introduction of such dual functions, one is able to obtain
bijective deformed functions, $\ln_{\bxi}$ and $\exp_{\bxi}$, which are called complete.
If $\mc{A}_{\bxi},\,\mc{B}^{(k)}_{\bxi}\subseteq\mb{R}$ are
parameter ranges related over a function
$d_k:\mc{A}_{\bxi}\ra\mc{B}^{(k)}_{\bxi}$ such that
$f_{\bxi}(x)=-f_{d_k(\bxi)}(1/x)$ and $h_{\bxi}(x)=1/h_{d_k(\bxi)}(-x)$,
then for $\bxi\in\mc{A}_{\bxi}\; (d_k(\bxi)\in\mc{B}^{(k)}_{\bxi})$, the complete
deformed functions are defined in the following domains and
codomains, $\ln_{\bxi}:\{f_{d_k(\bxi)}(x),\:x\in(0,1]\}\bigcup\{f_{\bxi}(x),\:x\in[1,\infty)\}$ $\lra$
$\{f_{d_k(\bxi)}(x)\in(-\infty,0]\}\bigcup\{f_{\bxi}(x)\in[0,\infty)\}$ and
$\exp_{\bxi}:\{h_{d_k(\bxi)}(x),\:x\in(-\infty,0]\}\bigcup\{h_{\bxi}(x),\:x\in[0,\infty)\}$ $\lra$ $\{h_{d_k(\bxi)}(x)\in(0,1]\}\bigcup\{h_{\bxi}(x)\in[1,\infty)\}$, with
$h_{\bxi}\equiv f^{-1}_{\bxi}$ and $h_{d_k(\bxi)}\equiv f^{-1}_{d_k(\bxi)}$.
Further, we
posited conditions that the generalized logarithmic and exponential
functions have to fulfill in order to preserve the main properties
of their ordinary counterparts e.g., the behavior at the boundary
points and the points of extrema. Through these conditions, we
determine the ranges $\mc{A}_{\bxi}$ and $\mc{B}^{(k)}_{\bxi}$.

Concerning the dual functions $d_k(\bxi)$, we formulated a criterion to
distinguish between deformed logarithmic/exponential structures
which are bijective in the original (co)domains, preserving the same
parameter for all values of the argument, and those whose
bijectivity is assured in the intervals presented above.

The application of our theoretical formalism to Tsallis
$q$-definitions led to the parameter ranges $\mc{A}_q=(0,1]$ and
$\mc{B}_q^{(1)}=[1,2)$ connected to one another through the dual function
$d(q)=2-q$. Moreover, writing the R\'enyi entropy in terms of
Tsallis deformed functions, we have obtained the interval of
concavity associated with R\'enyi entropy by the estimation of the
parameter range resulted from the criterion of completeness.

Similarly, we considered Kaniadakis generalized functions and
derived the parameter intervals $\mc{A}_\kappa=[0,1)$ and
$\mc{B}^{(1)}_\kappa=(-1,0]$ connected through the dual function
$d(\kappa)=-\kappa$. However, Kaniadakis generalizations satisfy, in
complete analogy to the ordinary functions, the identities
$f_{\kappa}(x)=-f_{\kappa}(1/x)$ and
$f^{-1}_{\kappa}(x)=1/f^{-1}_{\kappa}(-x)$, revealing their
$\kappa$-symmetric structure. Due to this property of Kaniadakis
generalized functions, the introduction of the dual function is
unnecessary. Indeed, we verify that the $\kappa$- and
$d(\kappa)$-structure of $f$ or $f^{-1}$ are identical. In this
case, the aforementioned deformed maps preserve the same
parameter values in their entire domain and codomain. The existence
of two parameter ranges $\mc{A}_\kappa$ and $\mc{B}^{(1)}_\kappa$ is not
in contradiction with the latter statement, since the
$\kappa$-symmetric generalizations give the same image whether
$\kappa\in\mc{A}_\kappa$ or $d(\kappa)\in\mc{B}^{(1)}_\kappa$.

Our final applications were Abe and Borges-Roditi deformed
functions. In each case, we were able to determine the respective
parameter intervals and the duality functions. Since the
Borges-Roditi generalization scheme includes all three deformed
logarithmic (and deformed exponential) definitions i.e., Tsallis,
Kaniadakis and Abe deformed functions, the compact form of the dual
function and the parameter intervals associated with it were found
to reduce to those of the Tsallis, Kaniadakis and Abe deformed
functions by identifying $\{\xi_1=1-q,\,\xi_2=0\}$,
$\{\xi_1=\kappa=-\xi_2\}$ and $\{\xi_1=1/q-1,\,\xi_2=q-1\}$,
respectively. The Borges-Roditi logarithmic family includes only one
possible complete and analytic structure in all points of its
domain, namely, when $\xi_2=-\xi_1$. This corresponds to the
Kaniadakis generalization scheme.

It is worth noting that the criterion of completeness developed here is not an alternative to the generalized sum or product rules. Although these generalized algebras may help preserve some properties of the ordinary exponential and logarithmic functions, they are nevertheless defined in terms of generalized functions. Therefore, they too suffer from the incompleteness of the deformed functions. Moreover, the criterion of completeness is far more general than the approach of generalized algebras, since the latter may drastically change from one generalization to the other. The criterion of completeness, on the other hand, is always based on the same procedure i.e., calculating the dual functions and the valid parameter intervals.

The current results unveil new perspectives on the consideration of
generalized logarithmic and exponential functions, thereby shedding
light on various mathematical aspects in need of revision and proper
treatment.

\section*{Acknowledgments}
\noindent TO acknowledges fruitful remarks from E.M.F. Curado, C. Tsallis, R.S. Wedemann and L. Lacasa. We thank U. Tirnakli for a careful reading of the manuscript and bringing Ref. [16] to our attention. GBB was supported by TUBITAK (Turkish Agency) under the Research Project number 108T013. TO was supported by CNPq (Brazilian Agency) under the Research Project number 505453/2008-8.


\end{document}